\documentclass{elsart}
\usepackage{natbib}
\usepackage{epsfig}
\begin{document}
\runauthor{Cicero, Caesar and Vergil}
\begin{frontmatter}
\title{A fresh look at neutral meson mixing }
\author{M.A. Gomshi Nobary, B. Mojaveri}
\thanks[]{E-mail: mnobary@razi.ac.ir}
\address{Department of Physics, Faculty of Science, Razi University, Kermanshah, Iran.}
\begin{abstract}
In this work we show that the existence of a complete
biorthonormal set of eigenvectors of the effective Hamiltonian
governing the time evolution of neutral meson system is a
necessary condition for diagonalizability of such a Hamiltonian.
We also study the possibility of probing the $CPT$ invariance by
observing the time dependence of cascade decays of type
$P^{\circ}(\overline{ P^{\circ}})\rightarrow
\{M_a,M_b\}X\rightarrow fX$ by employing such basis and exactly
determine the $CPT$ violation parameter by comparing the time
dependence of the cascade decays of tagged $P^{\circ}$ and tagged
$\overline{ P^{\circ}}$.
\end{abstract}
\end{frontmatter}

\section{Introduction}

In the Wigner-Weisskopf (W-W) approximation [1] the effective
Hamiltonian which describes the $P^{\circ} - \overline{
P^{\circ}}$ system is not Hermitian. Therefore the eigenkets of
this Hamiltonian are indistinguishable (unless the Hamiltonian is
normal, $[\hat H,\hat H^\dag]=0$). The reason is that for such a
system the orthogonality and completeness relations could not be
written in terms of its eigenkets. In the presence of $T$
violation in the $P^{\circ} - \overline{ P^{\circ}}$ system, we
are dealing with a non-hermitian Hamiltonian which is not normal.
Therefore we can use the principles of non-hermitian quantum
mechanics and reconsider the definition of diagonalizability of an
operator. We use the biorthonormal basis for this propose. Here we
emphasis that when a non-hermitian and non-normal operator is
encountered, use of a complete biorthonormal basis is in order
which reduces to orthogonal basis as soon as the operator is
considered hermitian and normal. Therefore we conclude that we may
use such a set of basis to describe the time evolution of neutral
mesons in the presence of $T$ violation.

As mentioned, in the presence of $T$ violation the eigenkets of a
non-hermition Hamiltonian does not satisfy the completeness and
orthogonality relations. Therefore the eigenkets of such a
Hamiltonian are not distinguishable. Due to this fact in writing
down the transition amplitudes for cascade decays
$P^{\circ}(\overline{ P^{\circ}})\rightarrow
\{M_a,M_b\}X\rightarrow fX$, we use the biorthonormal basis when
the intermediate states are eigenstates of $\hat H$.

We prove that the existance of a complete biorthonormal set of
eigenvectors of $\hat H$ is necessary condition for
diagonalizability of the effective Hamiltonian governing the time
evolution of the neutral meson systems and write down the spectral
form of the Hamiltonian operator with this basis in section 2. In
section 3 we discuss the time evolution of neutral meson system
and introduce the $T$ and $CPT$ violation complex parameters and
obtain the time evolution of flavor eigenkets. Finally in the last
section we study the possibility of probing $CPT$ invariance by
observation of the time dependence of the cascade decays of type
$P^{\circ}(\overline{ P^{\circ}})\rightarrow
\{M_a,M_b\}X\rightarrow fX$ by using the biorthonormal basis and
introduce new ratios of decay amplitudes and exactly determine the
$CPT$ violation parameter by comparing the time dependence of the
cascade decays of tagged $P^{\circ}$ and tagged $\overline{
P^{\circ}}$.

\section{Diagonalizability and the complete set of biorthonormal basis.}

A linear operator $\hat H$ acting in a separable Hilbert space and
having a discrete spectrum is diagonalizable if an only if there
are eigenvectors $|\psi_n\rangle$ of $\hat H$ and $|\phi_n\rangle$
of $\hat H^\dag$ that form a complete set of biorthonormal basis
of $\{\psi_n, \phi_n\}$, i.e. they satisfy

\begin{eqnarray}
\hat H |\psi_n\rangle=E_n |\psi_n\rangle,\quad \hat H^\dag |\phi_n
\rangle=E_n^* |\phi_n\rangle,
\end{eqnarray}
and
\begin{eqnarray}
\langle\psi_m|\phi_n\rangle=\delta_{mn},\quad \sum _n
|\psi_n\rangle\langle\phi_n|=\sum_n
|\phi_n\rangle\langle\psi_n|=1.
\end{eqnarray}

Where $n$ is the spectral label and $\dag$ and * denote the
adjoint and complex-conjugate respectively as usual. Moreover that
$\delta_{mn}$ is the Kronecker delta function and 1 represents the
identity operator.

Nowhere in this definition it is assume that the operator is
normal, i.e., $[{{\hat H}},{ \hat H}^\dag$]=0. A normal operator,
in finite dimensions with no extra conditions and in
infinite-dimensions with appropriate extra conditions, admit a
diagonal metrix representation in some orthogonal basis. This is
usually called {\it diagonalizability} by a unitary
transformation. In view of (1) and (2) The spectral form of $\hat
H$ and $\hat H^\dag$ may be written in the following form

\begin{eqnarray}
\hat H=\sum_n E_n|\psi_n\rangle\langle\phi_n|,\quad \hat
H^\dag=\sum_n E_n^*|\phi_n\rangle\langle\psi_n|.
\end{eqnarray}

In order to see the equivalence of the existence of a complete
biorthonormal set of eigenvectors of $\hat H$ and its
diagonalizability, we note that by definition a diagonalizable
Hamiltonian $\hat H$ satisfies $\hat {A}^{-1}\hat H \hat A=\hat
H_\circ$ for an invertible linear operator $\hat A$ and a diagonal
linear operator $\hat H_\circ$, i.e., there is an orthogonal basis
$\{|n\rangle\}$ in the Hilbert space and complex numbers $E_n$
such that $\hat H_\circ=\sum_n E_n|n\rangle\langle n |$. Then
letting $|\psi_n\rangle :=\hat A|n\rangle$ and $|\phi_n\rangle
:=(\hat {A}^{-1})^\dag|n\rangle$, we can easily check that
$\{|\psi_n\rangle ,|\phi_n\rangle \}$ is a complete biorthonormal
system for $\hat H$. The converse is also true, for if such a
system exists we may set $\hat A :=\sum_n |\psi_n\rangle\langle
n|$ for some orthogonal basis $\{|n\rangle\}$ and by using
equation (2) check that $\hat
{A}^{-1}=\sum_n|n\rangle\langle|\phi_n|$ and $\hat {A}^{-1}\hat
{H}\hat A=\hat H_\circ$, i.e., $\hat H$ is diagonalizable.

As long as $T$ is invariant (no violation), the effective
Hamiltonian is normal. In such a case the orthonormality relations
between the basis of $\hat H$ are valid and moreover that the
eigenket of $\hat H$ are discriminant and the biorthonormal basis
turn into orthonormal basis automatically and $|\psi_n\rangle$'s
are the same as $|\phi_n\rangle$'s.

\section{The time evolution of neutral meson system}

In the Wigner-Weisskoff (W-W) approximation, which we shall use
throughout, a beam of oscillating and decaying neutral meson
system is described in its rest frame by a two component wave
function

\begin{eqnarray}
|\psi(t)\rangle=\psi_1(t)|P^\circ\rangle+\psi_2(t)|\overline
{P^\circ}\rangle,
\end{eqnarray}

where $t$ is the proper time and $|P^\circ\rangle$ stands for
$K,\;D,\;B_d$ or $B_s$. The wave function evolves according to a
Schr\"{o}dinger like equation

\begin{eqnarray}
i\frac{d}{dt}\left (
\begin{array}{c}\psi_1(t)\\ \psi_2(t)
\end{array}\right )=\left (
\begin{array}{cc}H_{11}&H_{12}\\H_{21}&H_{22}
\end{array}\right )\left (
\begin{array}{c}\psi_1(t)\\ \psi_2(t)
\end{array}\right ).
\end{eqnarray}

The matrix $\hat H$ is usually written as $\hat H=\hat
M-i\hat\Gamma/2$. Where $\hat M=\hat M^\dag$, and
$\hat\Gamma=\hat\Gamma^\dag$ are $2\times 2$ matrices called the
mass and the decay matrices [1-7]. Decomposition of $\hat H$ reads
\begin{eqnarray}
\hat H=|P^\circ\rangle H_{11}\langle P^\circ|+|P^\circ\rangle
H_{12}\langle \overline{ P^\circ}|+|\overline{ P^\circ}\rangle
H_{21}\langle P^\circ|+|\overline{P^\circ}\rangle H_{22}\langle
\overline{P^\circ}|.
\end{eqnarray}

The flavor basis $\{|P^\circ\rangle, |\overline{P^\circ}\rangle\}$
satisfy orthogonality and completeness relations

\begin{eqnarray}
\langle P^\circ|\overline{P^\circ}\rangle&=&\langle
\overline{P^\circ}| P^\circ\rangle=0,\quad \langle P^\circ |
P^\circ\rangle=\langle \overline{ P^\circ} |\overline{
P^\circ}\rangle=1,\nonumber\\ |P^\circ\rangle \langle
P^\circ|&+&|\overline{P^\circ}\rangle \langle
\overline{P^\circ}|=1.
\end{eqnarray}

It is readily shown that $\hat H$ is not hermitian. If $[\hat H,
\hat H^\dag]\neq 0$ then, the orthogonality and completeness
relations for eigenstates of $\hat H$ are not satisfied, i.e., the
eigenstates of $\hat H$ are not discriminant states. Therefore we
cannot diagonalize $\hat H$ or write its spectral form though its
basis. To do this job we make the benefit of biortonormal basis.
Such basis could be set up for neutral meson system. indeed the
Hamiltonian is diagonalizable only with such set of basis.

According to section 2 the eigenvalues of $\hat H$ are denoted by
$\mu_a=m_a-i\Gamma_a/2$ and $\mu_b=m_b-i\Gamma_b/2$ corresponding
to the eigenvectors $|P_a\rangle$ and $|P_b\rangle$ respectively.
So that

\begin{eqnarray}
\hat H|P_a\rangle&=&\mu_a|P_a\rangle,\nonumber\\
\hat H|P_b\rangle&=&\mu_b|P_b\rangle.
\end{eqnarray}

We also denote the eigenvalues of $\hat H^\dag$ by
$\mu_a^*=m_a+i\Gamma_a/2$ and $\mu_b^*=m_b+i\Gamma_b/2$
corresponding to the eigenvectors $|\widetilde{P_a}\rangle$ and
$|\widetilde{P_b}\rangle$ respectively. So that

\begin{eqnarray}
\hat H^\dag|\widetilde{P_a}\rangle&=&\mu_a^*|\widetilde{P_a}\rangle , \nonumber\\
\hat
H^\dag|\widetilde{P_b}\rangle&=&\mu_b^*|\widetilde{P_b}\rangle .
\end{eqnarray}
It is not difficult to check that the set $\{
|{P_n}\rangle,|\widetilde{P_n}\rangle\}$, $n=a,b$, is a complete
biorthonormal system for $\hat H$ such that
\begin{eqnarray}
&&\langle P_a|\widetilde{P}_b\rangle=\langle \widetilde{P}_a|
P_b\rangle=0,\quad \langle P_a|\widetilde{P}_a\rangle=\langle
\widetilde{P}_b| P_b\rangle=1,\nonumber\\ \\
&&|P_a\rangle\langle\widetilde{P}_a|+|P_b\rangle\langle\widetilde{P}_b|=
|\widetilde{P}_a\rangle\langle{P}_a|+|\widetilde{P}_b\rangle\langle{P}_b|=1.
\end{eqnarray}

According to the definition given in the pervious section, $\hat
H$ could be diagonalized such that

\begin{eqnarray}
\chi^{-1}\hat H\chi=\left (
\begin{array}{cc}\mu_a&0\\ 0&\mu_b
\end{array}\right ),\quad \chi=\left (
\begin{array}{cc}p_a&q_a\\ p_b&-q_b
\end{array}\right ),
\end{eqnarray}
which means

\begin{eqnarray}
|P_a\rangle&=&p_a|P^\circ\rangle+q_a|\overline{P^\circ}\rangle ,\nonumber\\
|P_b\rangle&=&p_b|P^\circ\rangle-q_b|\overline{P^\circ}\rangle ,
\end{eqnarray}
and
\begin{eqnarray}
|\widetilde{P}_a\rangle=\frac{1}{p_aq_b+p_bq_a}\Bigl[q_b|P^\circ\rangle+
p_b|\overline{P^\circ}\rangle\Bigr],\nonumber\\
|\widetilde{P}_b\rangle=\frac{1}{p_aq_b+p_bq_a}\Bigl[q_a|P^\circ\rangle-
p_a|\overline{P^\circ}\rangle\Bigr].\nonumber\\
\end{eqnarray}

The signs in front of $q_a$ and $q_b$ in (13) and $p_a$ and $p_b$
in (14) are just a convention which may differ among different
authors, or even from one neutral meson system to another within
the same paper. Now we can write the spectral form of the
Hamiltonian $H$

\begin{eqnarray}
\hat H=\mu_a|{P}_a\rangle\langle
\widetilde{P_a}|+\mu_b|{P}_b\rangle\langle \widetilde{P_b}|.
\end{eqnarray}

The normalization conditions are

\begin{eqnarray}
|p_a|^2+|q_a|^2=|p_b|^2+|q_b|^2=1.
\end{eqnarray}

We find from equations (8) and (13) that

\begin{eqnarray}
\frac{q_a}{p_a}=\frac{\mu_a-H_{11}}{H_{12}}=\frac{H_{21}}{\mu_a-H_{22}},
\end{eqnarray}

\begin{eqnarray}
\frac{q_b}{p_b}=\frac{H_{11}-\mu_b}{H_{12}}=\frac{H_{21}}{H_{22}-\mu_b}.
\end{eqnarray}

By considering the effect of discrete symmetries on the matrix
elements of $\hat H$ we have

\begin{eqnarray}
CPT \;{\rm conservation}&\rightarrow& H_{11}=H_{22},\nonumber\\
T \;{\rm conservation}&\rightarrow& |H_{12}|=|H_{21}|,\nonumber\\
CP\; {\rm conservation}&\rightarrow& H_{11}=H_{22}\; {\rm and
}\;|H_{12}|=|H_{21}|.\nonumber
\end{eqnarray}

The above conditions suggest the dimensionless complex $CP$ and
$CPT$ parameter as [2]

\begin{eqnarray}
\theta\equiv\frac{\frac{q_a}{p_a}-\frac{q_b}{p_b}}{\frac{q_a}{p_a}+\frac{q_b}{p_b}}
=\frac{H_{22}-H_{11}}{\mu_a-\mu_b},
\end{eqnarray}
and the $CP$ and $T$ violation parameter as

\begin{eqnarray}
\delta\equiv\frac{|\frac{p_b}{q_b}|-|\frac{q_a}{p_a}|}{|\frac{p_b}{q_b}|+|\frac{p_a}{q_a}|}
=\frac{|H_{12}|-|H_{21}|}{|H_{12}|+|H_{21}|}.
\end{eqnarray}

It is convenient to introduce

\begin{eqnarray}
\frac{q}{p}=\sqrt{\frac{q_aq_b}{p_ap_b}}=\sqrt{\frac{H_{21}}{H_{12}}}.
\end{eqnarray}

If $CPT$ violation is absent from the mixing, then
$q/p={q_a}/{p_a}={q_b}/{p_b}$ and $\sqrt{1-\theta^2}=1$. In that
case one only needs to use ${q}/{p}$.

The time evolution of the neutral meson system is easily obtained
using the spectral form of the Hamiltonian $\hat H$

\begin{eqnarray}
e^{-i \hat H t}=e^{-i\mu_a t}|P_a\rangle\langle\widetilde{P}_a|+
e^{-i\mu_b t}|P_b\rangle\langle\widetilde{P}_b|.
\end{eqnarray}

Using the eqs. (13), (19) and (22) one finds at time $t$ for the
states $|P^\circ\rangle$ and $|\overline{ P^\circ}\rangle$ created
at time $t=0$,

\begin{eqnarray}
|P^\circ(t)\rangle&=&\Bigl[g_+(t)-\theta
g_-(t)\Bigr]|P^\circ\rangle+
\frac{q}{p}\sqrt{1-\theta^2}g_-(t)|\overline{ P^\circ}\rangle,\nonumber\\
\\ |\overline{
P^\circ}(t)\rangle&=&\Bigl[\frac{p}{q}\sqrt{1-\theta^2}g_-(t)\Bigr]|P^\circ\rangle+
\Bigl[g_+(t)+\theta g_-(t)\Bigr]|\overline{P^\circ}\rangle,
\end{eqnarray}

where

\begin{eqnarray}
g_{\pm}(t)=\frac{1}{2}\Bigl(e^{-i\mu_at}\pm e^{-i\mu_bt}\Bigr).
\end{eqnarray}

\section{Cascade decay and CPT violation}

It was conjectured by Azimov [7] that additional tests of $CPT$
invariance (violation) could be performed by looking for
$B_d^{\circ}(\overline{ B_d^{\circ}})\rightarrow J/\psi
K^{\circ}(\overline{ K^{\circ}})\rightarrow J/\psi f$ {\it cascade
decay}, involving the neutral $B_d$ and neutral kaon system in
succession. This idea has been followed by Dass and Sarman [8]. In
all these cases, an initial $B^\circ$ meson ( $B^\circ$ stands for
both $B_d^\circ$ and $B_s^\circ$) can only decay to one of the
kaon's flavor eigenstates. To the leading order in the Standard
model, the decays $B_d^{\circ}\rightarrow \overline{ K^{\circ}}+X$
and $B_s^{\circ}\rightarrow{ K^{\circ}}+X$ and respective $CP$
conjugate decays are forbidden [9], [10]. The possibility of
probing $CPT$ invariance by observing the time dependence of the
cascade decays of the type $B_d^{\circ}(\overline{
B_d^{\circ}})\rightarrow J/\psi K^{\circ}/\overline{
K^{\circ}}\rightarrow J/\psi f$ was investigated more recently
[11] by considering that in new physics and in higher order, one
cannot neglect the processes $B_d^{\circ}\rightarrow J/\psi
\overline{ K^{\circ}} $ and $\overline {B^{\circ}_d}\rightarrow
J/\psi{ K^{\circ}} $ when considering such a radical possibility
as $CPT$ violation.

In such cases there are two times and two $CPT$ violation
parameters. The time $t$ for $B_d^{\circ}$ meson to oscillate
before decaying into $J/\psi K^{\circ}$ and time $t'$ in which
$K^{\circ}$ oscillates before decaying into $f$. The $CPT$
violation parameter $\theta$ in the ($B_d^{\circ}-\overline
{B^{\circ}_d}$) meson mixing and the $CPT$ violation parameter
$\theta '$ in the ($K^{\circ}-\overline {K^{\circ}}$) meson
mixing. It is possible to determine $\theta$ by comparing the $t$
dependence of the cascade decays of tagged $P^\circ$ and tagged
$\overline P^\circ$. Indeed, $\theta$ is computed in much the same
way as from the time dependence of non-cascade decays [12], [13].
The parameter $\theta '$ cannot be determined, because it always
appears entangled with some undetermined ratios of decay
amplitudes.

We study the possibility of probing $CPT$ invariance by
observation of the time dependence of the cascade decay of the
type $B_d^{\circ}(\overline{ B^{\circ}_d})\rightarrow J/\psi
\{K_L,K_S\}\rightarrow J/\psi f$ by employing the complete
biorthonormal basis and introduce new ratios of decay amplitudes.
In such cascade decays there are two times and one $CPT$-violation
parameter. The time $t$ for the $B_d^{\circ}$ meson to oscillate
before decaying into $J/\psi\{K_S, K_L\}$ and time $t'$ in which
the decay into final state $J/\psi f$ takes place. Since $CP$ is
violated, there is no final state that can be obtained only from
$K_S$ ( or $K_L$) and not from $K_L$ ( or $K_S$). Therefore all
calculations must involve the full transition chain

\begin{eqnarray}
i\rightarrow J/\psi \{K_L,K_S\}\rightarrow J/\psi f.
\end{eqnarray}

We consider an experiment in which a tagged $B_d^{\circ}$ evolves
for time $t$ and decays into an intermediate state $J/\psi K_L$ or
$J/\psi K_S$, which after time $t'$ decays into the final state
$J/\psi f$. The amplitude for this process is

\begin{eqnarray}
e^{-i\mu_S t'}\langle f|\hat T|K_S\rangle\langle \widetilde{K}_S
J/\psi|\hat T|B_d^\circ (t)\rangle+e^{-i\mu_L t'}\langle f|\hat
T|K_L\rangle\langle \widetilde{K}_LJ/\psi|\hat T|B_d^\circ
(t)\rangle.
\end{eqnarray}

By using eqs. (23) and (24) (time evolution of flavor eigenstates)
it is easy to show that the above expression casts into the
following

\begin{eqnarray}
&&\frac{1}{2}\langle f|\hat T|K_S\rangle e^{-i\mu_S
t'}\Bigl[\bigl((1-\theta)e^{-i\mu_a t}+(1+\theta)e^{-i\mu_b
t}\bigr)\langle \widetilde{K}_SJ/\psi |\hat
T|B_d^\circ\rangle\nonumber\\
&&+\frac{q}{p}\sqrt{1-\theta^2}\Bigl(e^{-i\mu_a t}-e^{-i\mu_b
t}\Bigr)\langle \widetilde{K}_S J/\psi |\hat
T|\overline {B^\circ_d}\rangle\Bigr]\nonumber\\
&&+\frac{1}{2}\langle f|\hat T|K_L\rangle e^{-i\mu_L
t'}\Bigl[\bigl((1-\theta)e^{-i\mu_a t}+(1+\theta)e^{-i\mu_b
t}\bigr)\langle \widetilde{K}_LJ/\psi |\hat
T|B_d^\circ\rangle\nonumber\\
&&+\frac{q}{p}\sqrt{1-\theta^2}\Bigl(e^{-i\mu_a t}-e^{-i\mu_b
t}\Bigr)\langle \widetilde{K}_L J/\psi |\hat T|\overline
{B^\circ_d}\rangle\Bigr].
\end{eqnarray}

Now we introduce the four parameters

\begin{eqnarray}
\lambda= -\frac{q}{p}\frac{\langle \widetilde {K}_S J/\psi | \hat
T|\overline{ B^\circ}\rangle}{\langle \widetilde K_S J/\psi |\hat
T|{B^\circ}\rangle },
\end{eqnarray}

\begin{eqnarray}
y= -\frac{q}{p}\frac{\langle \widetilde{K}_L J/\psi |\hat
T|{B^\circ}\rangle}{\langle \widetilde K_S J/\psi|\hat
T|{B^\circ}\rangle },
\end{eqnarray}

\begin{eqnarray}
\overline{ y}=-\frac{\langle \widetilde{K}_L J/\psi |\hat
T|\overline{ B^\circ}\rangle}{\langle \widetilde K_S J/\psi|\hat
T|{B^\circ}\rangle },
\end{eqnarray}

and

\begin{eqnarray}
\eta_f=  \frac{\langle f |\hat T|K_L\rangle}{\langle f|\hat
T|K_S\rangle }.
\end{eqnarray}

Therefore we may write the amplitude as

\begin{eqnarray}
&& e^{-i\mu_S t'}\langle f|\hat T|K_S\rangle\langle
\widetilde{K}_SJ/\psi |\hat T|B_d^\circ (t)\rangle +e^{-i\mu_L
t'}\langle f|\hat T|K_L\rangle\langle \widetilde{K}_L J/\psi|\hat
T|B_d^\circ (t)\rangle\nonumber\\
&&\propto e^{-i\mu_S t'}\Bigl[e^{-i\mu_a t}+Re^{-i\mu_b
t}\Bigr]+Q\eta_f e^{-i\mu_L t'}\Bigl[e^{-i\mu_a t}+Se^{-i\mu_b
t}\Bigr],
\end{eqnarray}

where

\begin{eqnarray}
R=\frac{(1+\theta)+\lambda \sqrt{1-\theta ^2}}{(1-\theta)-\lambda
\sqrt{1-\theta ^2}},\quad S=\frac{\overline y(1+\theta)-y
\sqrt{1-\theta ^2}}{\overline y(1-\theta)+ y\sqrt{1-\theta ^2}}.
\end{eqnarray}

We next consider the analogous experiment with an initial
$\overline B_d^\circ$. The amplitude is

\begin{eqnarray}
&& e^{-i\mu_S t'}\langle f|\hat T|K_S\rangle\langle
\widetilde{K}_SJ/\psi |\hat T|\overline{ B^\circ_d }(t)\rangle
+e^{-i\mu_L t'}\langle f|\hat T|K_L\rangle\langle \widetilde{K}_L
J/\psi|\hat
T|\overline {B^\circ_d} (t)\rangle\nonumber\\
&&\propto e^{-i\mu_S t'}\Bigl[e^{-i\mu_a t}+\overline R e^{-i\mu_b
t}\Bigr]+Q\eta_f e^{-i\mu_L t'}\Bigl[e^{-i\mu_a t}+\overline
Se^{-i\mu_b t}\Bigr],
\end{eqnarray}

where

\begin{eqnarray}
\overline R=\frac{\lambda (1-\theta)+\sqrt{1-\theta ^2}}{\lambda
(1+\theta)-\sqrt{1-\theta ^2}},\quad \overline S=\frac{
y(1-\theta)-\overline y \sqrt{1-\theta ^2}}{y(1+\theta)+ \overline
y \sqrt{1-\theta ^2}}.
\end{eqnarray}

It is easily checked that

\begin{eqnarray}
\overline{R}&=&-\frac{1-\theta}{1+\theta} R,\nonumber\\
\overline{S}&=&-\frac{1-\theta}{1+\theta} S.\nonumber\\
\end{eqnarray}

If we use the first order approximation for small parameters,
i.e., the approximation of neglecting all products of $\theta$ and
$\lambda$, then we find

\begin{eqnarray}
{R}\simeq (1+\lambda)(2\theta+\lambda+1),
\end{eqnarray}

\begin{eqnarray}
S\simeq \Bigl(1-\frac{y}{\overline
y}\Bigr)\Bigl(2\theta-\frac{y}{\overline y}+1 \Bigr).
\end{eqnarray}

One concludes from eq. (37) that the $CPT$ violation in $B$ mixing
( the parameter $\theta$) can in principle be determined either
from the comparison of $R$ and $\overline R$, or from the
comparison of $S$ and $\overline S$. Indeed, $\overline R \neq -R$
and $\overline S \neq -S$ unequivocally indicate the presence of
$CPT$ violation in the mixing of the $B$-meson. One can measure
The $CPT$ violation in $B$-meson mixing by observation of the time
dependence of the tagged cascade decays.

\section{Conclusion}

In this work we introduced the system of complete set of
biorthonormal basis as a necessary condition for diagonalizability
of non-hermitian Hamiltonian (regardless of being normal or not
normal). We also noticed that in the presence of $T$ violation,
since $\hat H$ is not normal, the mass eigenstates do not satisfy
completeness and orthonormal relations and therefore are
indistinguishable and non-physical.

We also studied the possibility of $CPT$ invariance in the cascade
model of the type $P^{\circ}(\overline{ P^{\circ}})\rightarrow
\{P_a,P_b\}X\rightarrow fX$ by observation of the time dependence
of the process. In this case there are two time parameters of $t$
and $t'$ and one $CPT$ violation parameter. We showed that it is
possible to determine the $\theta$ parameter by comparison of time
dependence of the cascade models for $B^\circ$ and $\overline
B^\circ$.

We conclude by remarking that in the cascade model of
$P^{\circ}(\overline{ P^{\circ}})\rightarrow X (M^\circ /\overline
{M^\circ}) \rightarrow Xf$ there are two $CPT$ violation parameter
($\theta$ and $\theta '$) where $\theta$ is the $CPT$ violation
parameter in  $P^\circ - \overline P^\circ$ and $\theta '$ is the
$CPT$ violation parameter in  $M^\circ - \overline {M^\circ}$
system. In this case the parameter $\theta '$ in indeterminable
due to entanglement with other parameters [12]. However using the
biorthonormal system of basis for $P^{\circ}(\overline{
P^{\circ}})\rightarrow \{P_a,P_b\}X\rightarrow fX$, there is only
one $CPT$ violation parameter which is quite determinable.

\end{document}